\documentstyle{mn}
\newif\ifAMStwofonts
\ifoldfss
  \ifCUPmtlplainloaded \else
    \NewTextAlphabet{textbfit} {cmbxti10} {}
    \NewTextAlphabet{textbfss} {cmssbx10} {}
    \NewMathAlphabet{mathbfit} {cmbxti10} {} 
    \NewMathAlphabet{mathbfss} {cmssbx10} {} 
  \fi
  \ifAMStwofonts
    \ifCUPmtlplainloaded \else
      \NewSymbolFont{upmath} {eurm10}
      \NewSymbolFont{AMSa} {msam10}
      \NewMathSymbol{\upi}     {0}{upmath}{19}
      \NewMathSymbol{\umu}     {0}{upmath}{16}
      \NewMathSymbol{\upartial}{0}{upmath}{40}
      \NewMathSymbol{\leqslant}{3}{AMSa}{36}
      \NewMathSymbol{\geqslant}{3}{AMSa}{3E}
      \let\oldle=3D\le     \let\oldleq=3D\leq
      \let\oldge=3D\ge     \let\oldgeq=3D\geq
      \let\leq=3D\leqslant \let\le=3D\leqslant
      \let\geq=3D\geqslant \let\ge=3D\geqslant
    \fi
  \fi
\fi 

\ifnfssone
  \newmathalphabet{\mathit}
  \addtoversion{normal}{\mathit}{cmr}{m}{it}
  \addtoversion{bold}{\mathit}{cmr}{bx}{it}
  \newmathalphabet{\mathbfit} 
  \addtoversion{normal}{\mathbfit}{cmr}{bx}{it}
  \addtoversion{bold}{\mathbfit}{cmr}{bx}{it}
  \newmathalphabet{\mathbfss} 
  \addtoversion{normal}{\mathbfss}{cmss}{bx}{n}
  \addtoversion{bold}{\mathbfss}{cmss}{bx}{n}
  \ifAMStwofonts
    \ifCUPmtlplainloaded \else
      \UseAMStwoboldmath
      \makeatletter
      \new@mathgroup\upmath@group
      \define@mathgroup\mv@normal\upmath@group{eur}{m}{n}
      \define@mathgroup\mv@bold\upmath@group{eur}{b}{n}
      \edef\UPM{\hexnumber\upmath@group}
      \new@mathgroup\amsa@group
      \define@mathgroup\mv@normal\amsa@group{msa}{m}{n}
      \define@mathgroup\mv@bold\amsa@group{msa}{m}{n}
      \edef\AMSa{\hexnumber\amsa@group}
      \makeatother
      \mathchardef\upi=3D"0\UPM19
      \mathchardef\umu=3D"0\UPM16
      \mathchardef\upartial=3D"0\UPM40
      \mathchardef\leqslant=3D"3\AMSa36
      \mathchardef\geqslant=3D"3\AMSa3E
      \let\oldle=3D\le     \let\oldleq=3D\leq
      \let\oldge=3D\ge     \let\oldgeq=3D\geq
      \let\leq=3D\leqslant \let\le=3D\leqslant
      \let\geq=3D\geqslant \let\ge=3D\geqslant
    \fi
  \fi
\fi 

\ifnfsstwo
  \DeclareMathAlphabet{\mathbfit}{OT1}{cmr}{bx}{it}
  \SetMathAlphabet\mathbfit{bold}{OT1}{cmr}{bx}{it}
  \DeclareMathAlphabet{\mathbfss}{OT1}{cmss}{bx}{n}
  \SetMathAlphabet\mathbfss{bold}{OT1}{cmss}{bx}{n}
  \ifAMStwofonts
    \ifCUPmtlplainloaded \else
      \DeclareSymbolFont{UPM}{U}{eur}{m}{n}
      \SetSymbolFont{UPM}{bold}{U}{eur}{b}{n}
      \DeclareSymbolFont{AMSa}{U}{msa}{m}{n}
      \DeclareMathSymbol{\upi}{0}{UPM}{"19}
      \DeclareMathSymbol{\umu}{0}{UPM}{"16}
      \DeclareMathSymbol{\upartial}{0}{UPM}{"40}
      \DeclareMathSymbol{\leqslant}{3}{AMSa}{"36}
      \DeclareMathSymbol{\geqslant}{3}{AMSa}{"3E}
      \let\oldle=3D\le     \let\oldleq=3D\leq
      \let\oldge=3D\ge     \let\oldgeq=3D\geq
      \let\leq=3D\leqslant \let\le=3D\leqslant
      \let\geq=3D\geqslant \let\ge=3D\geqslant
    \fi
  \fi
\fi 

\ifCUPmtlplainloaded \else
  \ifAMStwofonts \else 
    \def\upi{\pi}
    \def\umu{\mu}
    \def\upartial{\partial}
  \fi
\fi

\title[Magnetic field amplification in CDM anisotropic
collapse]{Magnetic field amplification in CDM anisotropic
collapse}

\author[M. Bruni, R. Maartens and C.G. Tsagas]
       {Marco Bruni,$^1$\thanks{marco.bruni@port.ac.uk} Roy
       Maartens$^1$\thanks{roy.maartens@port.ac.uk} and Christos G.
       Tsagas$^{1,2}$\thanks{ctsagas@maths.uct.ac.za}\\
       ${}^1$ Institute of Cosmology and Gravitation,
       Portsmouth University, Portsmouth~PO1~2EG, Britain\\
       ${}^2$ Department of Mathematics and Applied Mathematics,
       University of Cape Town, Rondebosch 7701, South Africa}
\date{\today}

\pagerange{\pageref{firstpage}--\pageref{lastpage}} \pubyear{2001}

\begin{document}

\maketitle

\label{firstpage}

\begin{abstract}
We use the Zel'dovich approximation to analyse the amplification
of magnetic fields in gravitational collapse of cold dark matter
during the mildly nonlinear regime, and identify two key features.
First, the anisotropy of collapse effectively eliminates one of
the magnetic components, confining the field in the pancake plane.
Second, in agreement with recent numerical simulations, we find
that the shear anisotropy can amplify the magnetic field well
beyond the isotropic case. Our results suggest that the magnetic
strengths observed in spiral and disk galaxies today might have
originated from seeds considerably weaker than previous estimates.
\end{abstract}


\section{Introduction}
Large scale magnetic fields, with strengths
$10^{-7}$--$10^{-5}$~G, have been observed in spiral and disc
galaxies, galaxy clusters and high redshift condensations
[see~\cite{K,HW} and references therein]. The most promising
explanation for the large scale galactic fields has been the
dynamo mechanism, with the required seeds coming either from local
astrophysical processes, such as battery effects, or from
primordial magnetogenesis [for recent reviews, see~\cite{GR,Wid}].
The linear evolution of large scale magnetic fields and their
implications for structure formation has been studied by several
authors [see, e.g.,~\cite{RR,W,PE,EF,TB,BS,TM}]. Certain aspects
of the mildly nonlinear clustering can be analysed in spherical
symmetry, but this approximation inevitably breaks down as the
collapse proceeds and any initially small anisotropies take over.
When magnetic fields are involved, the need to incorporate
anisotropic effects is particularly important, as the fields are
themselves generically anisotropic sources. Numerical simulations
of magnetic field evolution in galaxy clusters suggest that
anisotropies lead to additional amplification of the field. Tidal
effects during mergers, for example, increase the magnitude of the
magnetic field~\cite{RSB}. Shear flows in galaxy clusters can also
amplify the field beyond the limits of spherical
compression~\cite{DBL1,DBL2}. The latter simulations also suggest
that the final magnetic configuration is effectively independent
of the field's initial set up or of the presence of a cosmological
constant.

Here we use the Zel'dovich approximation to look analytically into
the effect of gravitational anisotropy on seed magnetic fields in
the mildly nonlinear regime. The anisotropic collapse is driven by
cold dark matter (CDM). A previous analysis~\cite{ZRS}, where the
matter is purely baryonic, concluded that the seeds must be
negligibly small in order to avoid the growth of magnetic fields
to levels which prevent pancake formation. By contrast, this
strict constraint is avoided when CDM dominates, since the
magnetic field couples only gravitationally with the CDM. The
field is frozen into the baryon fluid, and baryons are dragged by
the CDM gravitational field. The baryons are in a different state
of motion to the CDM and, unlike the CDM, they feel the magnetic
backreaction. As a first approximation, however, we will ignore
the magnetic backreaction and the relative motion of the baryons,
effectively considering a single fluid. This approximation also
maintains the acceleration-free and irrotational nature of the
motion, which are key ingredients of the Zel'dovich approximation.
Our approach may be seen as a qualitative starting point for a
more detailed analysis.

We begin in Section 2 with a dynamical system description of the
Zel'dovich approximation, which directly shows that, in a generic
collapse, pancakes are the (local) attractors~\cite{B}. In Section
3 we consider the dynamics of the magnetic field as it collapses
with the matter. The magnetized dynamical system is
five-dimensional. Pancakes are still the attractors, with the
magnetic field squeezed in the pancake plane. Note that, as the
galaxy is formed, tidal forces are generally expected to change
the orientation of the galactic plane relative to the pancake.
Nevertheless, the confinement of the field in the pancake plane
that we find here, is qualitatively consistent with magnetic field
observations in numerous spiral and disc galaxies. We also provide
some quantitative results by relating the growth of the field to
that of the matter density contrast. When comparing the shearing
to the isotropic collapse, we find that anisotropy can lead to an
appreciable increase in the amplification of the magnetic field.
These analytical results are in qualitative agreement with those
of the earlier mentioned numerical simulations. We interpret them
as an indication that the magnetic strengths observed in numerous
galaxies and galaxy clusters today could have resulted from seeds
considerably weaker than previous estimates.

\section{Zel'dovich dynamics as a 2-d system}
The Zel'dovich approximation~\cite{Z} [see also~\cite{SZ,P} for a
review and~\cite{buchert,mataa,matab,ET} for an approach similar
to ours] arises from a simple ansatz, which extrapolates to the
nonlinear regime a well known result of the linear perturbation
theory. In the Eulerian frame (${\bf x}, t$), which is comoving
with the background expansion, one defines the rescaled peculiar
velocity field ${\bf \tilde{u}}={\rm d} {\bf x}/{\rm d} a$, where
$a=a(t)$ is the background scale factor. Then, to linear order and
ignoring decaying modes, ${\bf \tilde{u}}=-\nabla_{\bf
x}\tilde{\varphi}$, where $\tilde{\varphi}=2\varphi/3H^2_0a_0^3$
is the rescaled peculiar gravitational potential and $H=\dot a/a$.
The Zel'dovich approximation uses this linear result in the
rescaled nonlinear continuity, Euler and Poisson equations, which
are
\begin{eqnarray}
\frac{\partial\delta}{\partial a}&=& -\nabla_{\bf
x}\cdot(\delta\tilde{\bf u})- \nabla_{\bf x}\tilde{\bf u}\,,
\label{eq:db}\\  \frac{\partial\tilde{\bf u}}{\partial a}&=&
-\tilde{\bf u}\cdot\nabla_{\bf x}\tilde{\bf u}-
\frac{3}{2a}\tilde{\bf u}- \frac{3}{2a}\nabla_{\bf
x}\tilde{\varphi}\,,  \label{eq:ub}\\ \nabla_{\bf
x}^2\tilde{\varphi}&=&\frac{1}{a}\delta\,,  \label{poisson}
\end{eqnarray}
where $\delta=\delta\rho/\rho_{\rm b}$ is the density contrast.
Assuming irrotational motion, we define the peculiar expansion and
the peculiar shear by $\tilde{\Theta}=\nabla_{\bf x}\cdot
\tilde{\bf u}$ and $\tilde{\sigma}_{ij}=\partial_{(j}
\tilde{u}_{i)} -{1\over3}\tilde{\Theta}\delta_{ij}$ respectively
($\partial_i\equiv\partial/\partial x_i$). Then, using convective
derivatives, Eqs.~(\ref{eq:db}) and (\ref{eq:ub}) give
\begin{eqnarray}
\frac{{\rm d}\delta}{{\rm d}a }&=&-(1+\delta)\tilde{\Theta}\,,
\label{cont}\\ \frac{{\rm d}\tilde{\Theta}}{{\rm d}a}&=&
-{1\over3}\tilde{\Theta}^2 -2\tilde{\sigma}^2-
\frac{3}{2a}(\tilde{\Theta}+\nabla_{\bf x}^2\tilde{\varphi})\,,
\label{eq:ray}\\ \frac{{\rm d}\tilde{\sigma}_{ij}}{{\rm d}a}&=&
{2\over3}\tilde{\sigma}^2\delta_{ij}-
{2\over3}\tilde{\Theta}\tilde{\sigma}_{ij}-
\tilde{\sigma}_{ik}\tilde{\sigma}_{kj}-
\frac{3}{2a}(\tilde{\sigma}_{ij} +E_{ij})\,,  \label{eq:shear}
\end{eqnarray}
where $E_{ij}=\partial_j\partial_i\tilde{\varphi}
-(\partial^2\tilde{\varphi}/3)\delta_{ij}$ is the Newtonian
traceless tidal field and
$2\tilde{\sigma}^2=\tilde{\sigma}_{ij}\tilde{\sigma}_{ij}$. One
can then substitute $\partial^2\tilde{\varphi}$ from
(\ref{poisson}) into (\ref{eq:ray}). In the linear, or in the
shear-free case, Eqs.~(\ref{cont}) and (\ref{eq:ray}) form a local
system describing the evolution of the fluid along its flow lines.
In the general case, the presence of the tidal term $E_{ij}$ in
(\ref{eq:shear}) and the lack of a corresponding evolution
equation show that the above system is non-local and not closed.
It also emphasises the fact that in Newtonian gravity, as opposed
to general relativity, one cannot consider a purely initial value
problem. Instead, as a consequence of action at a distance, one
necessarily needs boundary conditions~\cite{BG}.

The Zel'dovich ansatz,
\begin{equation}
\tilde{\bf u}=-\nabla_{\bf x}\tilde{\varphi}\,\hspace{3mm}
\Rightarrow \hspace{3mm} \tilde{\Theta}=-\nabla_{\bf
x}^2\tilde{\varphi}\,, \hspace{3mm} \tilde{\sigma}_{ij}=-
E_{ij}\,, \label{ansatz}
\end{equation}
implies that the parentheses in (\ref{eq:ray}) and
(\ref{eq:shear}) vanish, leading to a local system of ordinary
differential equations by eliminating the dependence on $\delta$
and $E_{ij}$. Since the shear and tide matrices now commute, the
above system can be written in the shear-tide eigenframe and
reduces to three equations, one for $\tilde{\Theta}$ and two for
the independent shear eigenvalues $\tilde{\sigma}_1$,
$\tilde{\sigma}_2$. It is then straightforward to verify that the
solutions of the reduced equations that follow from the Zel'dovich
ansatz are
\begin{eqnarray}
\tilde{\Theta}^{\rm zel}&=&\sum_{i=1}^{3}\,
\frac{\lambda_i}{1+a\lambda_i}\,,  \label{expZ}\\
\tilde{\sigma}_i^{\rm zel}&=&\frac{\lambda_i}{1+a\lambda_i}-
{1\over3}\,\tilde{\Theta}^{\rm zel}\,,  \label{shearZ}\\
{\delta}^{\rm zel}&=&\prod_{i=1}^{3}\,\frac{1}{1+a\lambda_i}- 1\,,
\label{deltaZ}
\end{eqnarray}
where $\lambda_i$ are the three eigenvalues of the initial tidal
field~\cite{mataa}. In particular, $\delta^{\rm zel}$ is the
solution of the continuity equation (\ref{cont}), when
$\tilde{\Theta}$ is given by (\ref{expZ}). On the other hand, if
$\delta^{\rm dyn}$ is the density contrast used in the Poisson
equation, then from (\ref{poisson}) and (\ref{ansatz}) one gets
$\delta^{\rm dyn}=-a\tilde{\Theta}^{\rm zel}\not=\delta^{\rm zel}$
as a consequence of the approximation~\cite{nusser}. Note that a
negative eigenvalue $\lambda_i$ corresponds to collapse along the
associated shear eigen-direction. Also, \mbox{1-dimensional}
planar pancakes are solutions corresponding to two vanishing
eigenvalues. For example, $\lambda_1=0=\lambda_2$ and
$\lambda_3<0$ describes \mbox{1-dimensional} collapse in the third
eigen-direction. As is well known, this is not only a solution of
the simplified local dynamics that follows from the Zel'dovich
ansatz, but also an exact solution of the full
system~(\ref{poisson})-(\ref{eq:shear}). In general one expects at
least one of the $\lambda_i$ to be negative and smaller than the
other two, and on this basis the generic solution should to tend
to a pancake.

The existence of pancake attractors can be confirmed by a
dynamical system approach~\cite{B}. Dropping the superscript
``zel'', we define the new time variable
\begin{equation}
\tau=-\int\tilde{\Theta}da\,,  \label{eq:tau}
\end{equation}
and the dimensionless dynamical variables
\begin{equation}
\Sigma_+={{3\over2}}(\Sigma_1+\Sigma_2)\,, \hspace{10mm}
\Sigma_-={{\sqrt{3}\over2}}(\Sigma_1-\Sigma_2)\,,  \label{eq:vars}
\end{equation}
where $\Sigma_i=\tilde{\sigma}_i/\tilde{\Theta}$, to arrive at the
system
\begin{eqnarray}
\frac{{\rm d}\Sigma_{+}}{{\rm d}\tau}&=&
{1\over3}\left(1-2\Sigma_+\right)
\left[\Sigma_{+}\left(\Sigma_{+}+1\right)+\Sigma_{-}^2\right]\,,
\label{eq:s1}\\ \frac{{\rm d}\Sigma_{-}}{{\rm d}\tau}&=& {1\over3}
\left[1-2\left(\Sigma_{+}^2+\Sigma_{-}^2\right)+2\Sigma_{+}\right]
\Sigma_{-}\,. \label{eq:s2}
\end{eqnarray}
Equation~(\ref{eq:ray}) has now decoupled and the problem has been
reduced to the study of the planar dynamical system (\ref{eq:s1}),
(\ref{eq:s2}) depicted in Fig.~1. The evolution of the shape of a
generic fluid element is given by the time dependence of its axes
$\ell_i$. In the shear eigenframe~\cite{E},
\begin{equation}
\frac{{\rm d}\ell_i}{{\rm d}a}=
\left({1\over3}+\Sigma_i\right)\tilde{\Theta}\ell_i\,. \label{ell}
\end{equation}
One-dimensional pancakes are locally axisymmetric solutions of the
planar system with two of the $\Sigma_i$ equal to $-1/3$, so that
two of the $\ell_i$ are constants. Generic pancakes on the other
hand, correspond to solutions that are asymptotic to one of the
three 1-dimensional pancakes (one for each shear eigen-direction).
In this case two of the $\ell_i$ tend to constants (i.e.~two of
the $\Sigma_i$ approach $-1/3$). Exact 2-dimensional ``filaments"
are solutions with two $\Sigma_i$ equal to $1/6$
(e.g.~$\Sigma_-=0$ and $\Sigma_+=1/2$).

The system (\ref{eq:s1}), (\ref{eq:s2}) admits two sets of
invariant sub-manifolds, corresponding to straight lines in the
$\Sigma_+,\Sigma_-$ plane of Fig.~1. The two sets of lines in
Fig.~1 are: (i) those that form the central triangle
(e.g.~$\Sigma_3=-1/3$, $\Sigma_+=1/2$), representing trajectories
with one of the $\Sigma_i=-1/3$; (ii) those that bisect the vertex
angles (e.g.~the $\Sigma_+$ axis, $\Sigma_-=0$) represent locally
axisymmetric solutions with two equal $\Sigma_i$. The vertices of
the triangle are the stationary points of (\ref{eq:s1}),
(\ref{eq:s2}) that correspond to exact 1-dimensional pancakes.
Where the first set of lines intersects the second, we have
stationary points representing exact filamentary solutions
(e.g.~$\Sigma_-=0$, $\Sigma_+=1/2$). Finally the centre, where the
three bisecting lines meet, corresponds to spherical collapse. It
is clear from Fig.~1 that depending on the initial conditions, a
generic fluid element collapses to one of the three pancakes,
while the three filaments are unstable. Indeed, looking at the
eigenvalues $s_\pm$ of the Jacobian of (\ref{eq:s1}),
(\ref{eq:s2}), one finds that the pancakes are asymptotically
stable nodes, while the filaments are saddle points, and the
spherical solution is an unstable node (see Table 1). This proves
that, once collapse sets in, pancakes are the attractors of the
generic trajectories.

\begin{table}
\caption{The fixed points in the $\Sigma_+$, $\Sigma_-$ plane. The
second and third columns give a typical example in the case
$\Sigma_1=\Sigma_2$ on the line $\Sigma_-=0$, the fourth and fifth
give the eigenvalues of the Jacobian of the dynamical system
(\ref{eq:s1}), (\ref{eq:s2}) at that fixed point.}
\vspace{0.5truecm}
\begin{tabular}{|l|r|r|r|r|l|} \hline
Fixed Point & $\Sigma_+$ & $\Sigma_-$ & $s_+$ & $s_-$ &  Stability
\\[1.5truemm]\hline
pancakes  \rule{0cm}{5truemm} & $-1$ & $0$ & $-1$ & $-1$ &
\parbox{2truecm}{asymptotically stable node}
\\[1.5truemm]\hline
filaments & $\frac{1}{2}$ & $0$ & $-\frac{1}{2}$ & $\frac{1}{2}$ &
saddle
\\[1.5truemm]\hline
spherical & $0$ & $0$ & $\frac{1}{3}$ & $\frac{1}{3}$ & unstable
node
\\[1.5truemm]\hline
\end{tabular}\vspace{0.5truecm}
\end{table}

\begin{figure}
\centering\vspace{5.5cm}\includegraphics{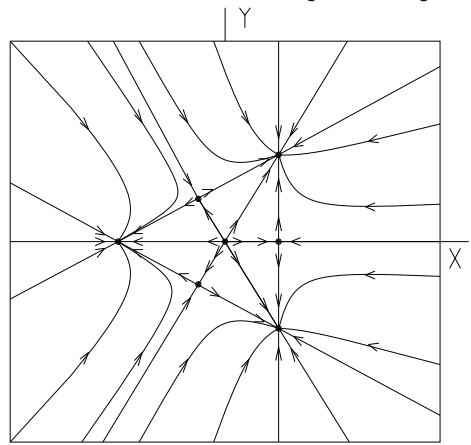} \caption{Phase plane
$\Sigma_{+}\equiv X$, $\Sigma_{-}\equiv Y$. The lines forming the
central triangle correspond to one of the $\Sigma_i=-1/3$. The
three pancakes are located at the intersections of these lines.
The lines bisecting the vertices correspond to two equal
$\Sigma_i$.} \label{fig:1}
\end{figure}

\section{Shear amplified magnetic fields}
We now analyse the effects of mildly nonlinear clustering on
magnetic fields frozen into a collapsing protocloud that is
falling into a CDM potential well. We neglect magnetic
backreaction on the fluid (effectively we consider a force-free
field), as well as the velocity of the baryons relative to the
CDM. In comoving coordinates the magnetic induction equation reads
\begin{equation}
\frac{{\rm d}{\bf B}}{{\rm d}t}=-2H{\bf B}+ \frac{1}{a}{\bf
B}\cdot\nabla_{\bf x}{\bf u}- \frac{1}{a}\nabla_{\bf x}{\bf
u}\cdot{\bf B}\,. \label{cIE}
\end{equation}
Introducing the rescaled magnetic field vector
$\tilde{B}_i=a^2B_i$, and using the shear eigenframe, this gives
\begin{eqnarray}
\frac{{\rm d}\tilde{B}_1}{{\rm d}\tau}&=&
-{1\over3}\left(\Sigma_{+}+\sqrt{3}\,\Sigma_{-}-2\right)\tilde{B}_1\,,
\label{B'1*}\\ \frac{{\rm d}\tilde{B}_2}{{\rm d}\tau}&=&
-{1\over3}\left(\Sigma_{+}-\sqrt{3}\,\Sigma_{-}-2\right)\tilde{B}_2\,,
\label{B'2*}\\ \frac{{\rm d}\tilde{B}_3}{{\rm d}\tau}&=&
{2\over3}\left(\Sigma_{+}+1\right)\tilde{B}_3\,.  \label{B'3*}
\end{eqnarray}

A pancake solution, with collapse along the third shear
eigen-direction, is characterised by the asymptotic values
$\Sigma_+=-1$ and $\Sigma_-=0$, so that
\begin{equation}
B_{1,2}\,\propto\,a^{-2}e^{\tau}
\,\propto\,a^{-2}\exp(-\int\tilde{\Theta}da),\, \hspace{5mm}
B_3\propto a^{-2}\,. \label{B1,2}
\end{equation}
In the above $a^{-2}$ gives the dilution due to the expansion,
$e^{\tau}$ the increase caused by the collapse and $\tau$ provides
a measure of the collapse timescale. Thus, we expect the magnetic
strength to increase as the field collapses with the fluid and
gets confined in the pancake plane. This qualitative result is
consistent with the pattern of the magnetic fields observed in
spiral and disc galaxies.

To obtain a more quantitative result we need to solve the magnetic
evolution equations (\ref{B'1*})-(\ref{B'3*}). Using expressions
(\ref{expZ}) and (\ref{shearZ}) we arrive at
\begin{eqnarray}
\frac{B_2}{B_{1_0}}&=&\left[\frac{(1+a_0\lambda_2)(1+a_0\lambda_3)}
{(1+a\lambda_2)(1+a\lambda_3)}\right]
\left(\frac{a_0}{a}\right)^2\,, \label{B1}\\
\frac{B_2}{B_{2_0}}&=&
\left[\frac{(1+a_0\lambda_1)(1+a_0\lambda_3)}
{(1+a\lambda_1)(1+a\lambda_3)}\right]
\left(\frac{a_0}{a}\right)^2\,, \label{B2}\\
\frac{B_3}{B_{3_0}}&=&
\left[\frac{(1+a_0\lambda_1)(1+a_0\lambda_2)}
{(1+a\lambda_1)(1+a\lambda_2)}\right]
\left(\frac{a_0}{a}\right)^2\,, \label{B3}
\end{eqnarray}
where $a_0$, $B_{i_0}$ are initial values. In 1-dimensional planar
collapse along the 3rd eigen-direction ($\lambda_1=0=\lambda_2$
and $\lambda_3<0$), the pancake singularity is reached as
$a\rightarrow-1/\lambda_3$. In that case $B_3$ decays, whereas
$B_1$, $B_2$ increase without bound (neglecting backreaction).

We now compare with spherical collapse, limiting to the Zel'dovich
case which is given by three equal eigenvalues. Then, any one of
(\ref{B1})-(\ref{B3}) reduces to
\begin{equation}
\frac{B^{\rm sph}}{B_0}=
\left(\frac{1+a_0\lambda}{1+a\lambda}\right)^2
\left(\frac{a_0}{a}\right)^2\,,  \label{SBi}
\end{equation}
and all the magnetic components diverge as we approach the point
singularity, $a\rightarrow-1/\lambda$. For comparison, we assume
equal initial values $B_{i_0}=B_0$, which leads to the ratio
\begin{equation}
\frac{B_1}{B^{\rm sph}}=
\frac{(1+a_0\lambda_2)(1+a_0\lambda_3)(1+a\lambda)^2}
{(1+a\lambda_2)(1+a\lambda_3)(1+a_0\lambda)^2}\,,
\label{Anis-Sph1}
\end{equation}
with an analogous relation for $B_2/B^{\rm sph}$. This result
shows that the relative growth of the field depends on the
eigenvalues corresponding to the two alternative patterns of
collapse. When $\lambda_3<\lambda<0$, or equivalently when
$0<-1/\lambda_3<-1/\lambda$, the anisotropy dominates the
collapse. Thus $a\rightarrow-1/\lambda_3$ before
$a\rightarrow-1/\lambda$, and therefore $B_1,\,B_2\gg B^{\rm
sph}$. In other words, anisotropy can lead to a considerable
increase in the strength of the magnetic field.

Conditions such as $\lambda_3<\lambda$ arise naturally whenever we
compare pancake to spherical collapse. Indeed, consider pure
pancake collapse along the 3rd shear eigen-direction
($\lambda_1=0=\lambda_2$). Then, from (\ref{expZ}) we have
\begin{equation}\label{r1}
\lambda_3={\tilde{\Theta}_0\over 1-a_0\tilde{\Theta}_0}\,.
\end{equation}
For spherical collapse on the other hand, Eq.~(\ref{expZ}) gives
\begin{equation}\label{r2}
\lambda={\tilde{\Theta}_0\over 3-a_0\tilde{\Theta}_0}\,,
\end{equation}
so that $\lambda_3<\lambda$. Accordingly, one should expect
stronger amplification for magnetic fields frozen into
anisotropic, rather than isotropic, collapse. Note that we have
implicitly assumed that, initially, the value of $\tilde{\Theta}$
is the same both for the pancake and for the spherical collapse.

\begin{figure}
\centering\vspace{6cm}\includegraphics{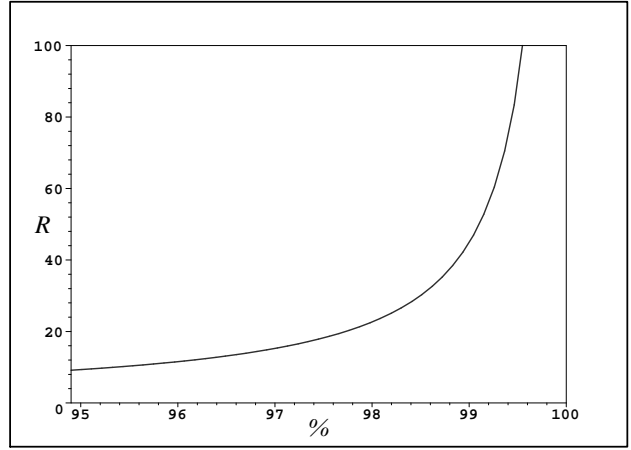}\caption{Typical growth of the ratio
$R={B_{1,2}}/{B^{\rm sph}}$, Eq.~(\ref{pan-Sph}), during the late
stages of the collapse. The horizontal axis measures the
percentage of the total pancake collapse according to
$(a-a_0)/a_0\times100\%$ ($a_0\leq a\leq2a_0$ for $\delta_0=1$).}
\label{fig:2}
\end{figure}

An alternative way of estimating the relative amplification is via
the initial density contrast. For $\lambda_1=0=\lambda_2$,
Eq.~(\ref{Anis-Sph1}) leads to
\begin{equation}
\frac{B_1}{B^{\rm sph}}=\frac{(1+a_0\lambda_3)(1+a\lambda)^2}
{(1+a\lambda_3)(1+a_0\lambda)^2}\,,  \label{purePan-Sph}
\end{equation}
with the same relation for $B_2/B^{\rm sph}$. Substituting for
$\lambda$, $\lambda_3$ from~(\ref{r1}) and (\ref{r2}) we arrive at
the expression
\begin{equation}
\frac{B_{1,2}}{B^{\rm
sph}}=\frac{[1-\delta_0/(1+\delta_0)][1-a\delta_0/a_0(3+\delta_0)]^2}
{[1-a\delta_0/a_0(1+\delta_0)][1-\delta_0/(3+\delta_0)]^2}\geq1\,,
\label{pan-Sph}
\end{equation}
where the equality on the right hand side holds at $a=a_0$. The
above directly relates the ratio $B_{1,2}/B^{\rm sph}$ to the
overdensity $\delta_0$ at the onset of the nonlinear regime. In
deriving Eq.~(\ref{pan-Sph}) we have also used the relation
$\delta_0=-a_0\tilde{\Theta}_0$, which arises directly from the
Zeldovich ansatz, Eq.~(\ref{ansatz}). According to
(\ref{pan-Sph}), the ratio $B_1/B^{\rm sph}$ keeps increasing as
the collapse proceeds towards the pancake singularity, namely as
$a\rightarrow a_0(1+\delta_0^{-1})$. At that point $B_{1,2}/B^{\rm
sph}$ diverges. Note that, for the same initial $\tilde{\Theta}$,
the isotropic collapse reaches the singularity at
$a=a_0(3+\delta_0^{-1})$, that is later than the pancake collapse.
On using Eq.~(\ref{pan-Sph}) we can calculate the relative
amplification of the magnetic field, for a given $\delta_0$, at
any time prior to the pancake singularity. To obtain a numerical
estimate consider the following example. Assume that $\delta_0=1$
at the onset of the nonlinear regime. This means that the pancake
singularity is reached at $a=2a_0$, whereas the point singularity
at $a=4a_0$. Then, the anisotropically collapsing magnetic field
is one order of magnitude stronger than an isotropically
collapsing one (i.e.~$B_{1,2}\simeq10B^{\rm sph}$) by the time
$a=1.95a_0$ (i.e.~when we have undergone 95\% of the total pancake
collapse). This means that the effect of the anisotropy on the
amplification, though weak early on, increases fast as we approach
the final stages of the collapse (see Fig.~\ref{fig:2}). Then, the
presence of shear can strengthen the magnetic field well beyond
the limits of simple isotropic compression. In other words, as the
field is dragged deep into the potential wells of the CDM, one
should expect an appreciable increase in the magnetic strength due
to shearing effects alone.

Although Eq.~(\ref{pan-Sph}) has been derived in the idealized
case of a 1-dimensional pancake, with the spherical case treated
via the Zel'dovich approximation, we expect it to be approximately
true in more realistic situations as well. This reasonable
expectation is further strengthened by the fact that our
analytical estimates are in good agreement with those obtained by
numerical simulations [see~\cite{DBL1}]. These simulations also
argue for an extra order of magnitude increase in the strength of
the magnetic seed field, during the final stages of the collapse,
because of shearing effects.

Finally, we can gain further insight into the relative
amplification of anistropically versus isotropically collapsing
magnetic fields by rewriting (\ref{purePan-Sph}) as
\begin{equation}
\frac{B_1}{B^{\rm sph}}=\frac{\ell_1}{\ell_3}\left(\frac{\ell^{\rm
sph}}{\ell^{\rm sph}_0}\right)^2=
\frac{\ell_1}{\ell_3}\left(\frac{2}{3}
+\frac{\ell_3}{3\ell_1}\right)^2\,, \label{eccentr}
\end{equation}
where we have also used Eqs.~(\ref{ell}), (\ref{r1}), (\ref{r2}),
and linear initial conditions for two initially spherical fluid
elements. This relation shows that the ratio ${B_1}/{B^{\rm sph}}$
grows almost linearly with the eccentricity of the pancake
$\ell_1/\ell_3$, independent of initial conditions.

\section{Conclusions}
Protogalactic clouds do not collapse isotropically: in any
realistic situation one expects small anisotropies in the initial
velocity distribution, which will be amplified as CDM collapse
progresses. We used the Zel'dovich approximation to investigate
the effects of such gravitational anisotropy on the evolution of a
seed magnetic field frozen into the baryon fluid. We considered
the mildly nonlinear regime and ignored the magnetic backreaction
on the baryons as well as the relative velocity between baryons
and CDM. The CDM dominates the collapse and the gravitational
anisotropy that amplifies the magnetic field. Our qualitative
analysis shows that, as a generic result of the anisotropy of the
collapse, the field effectively ``loses'' one of its components
and is confined in the plane of the pancake. Although tidal forces
are generally expected to change the orientation of the galactic
plane relative to the pancake, our qualitative picture is
consistent with magnetic field observations in numerous spiral and
disk galaxies. More quantitatively, our results show a much more
efficient magnetic amplification in a shearing rather than in a
shear-free collapse. This analytical result agrees with numerical
simulations of shear and tidal effects on the evolution of
magnetic fields in galaxies and galaxy clusters. Taken together,
these results indicate that, when the anisotropic effects of CDM
collapse are taken into account, the magnetic strengths observed
in galaxies today could have originated from seeds considerably
weaker than previous estimates.

\section*{Acknowledgments}
We thank John Barrow, Thanu Padmanabhan and particularly Kandu
Subramanian for helpful comments. MB thanks the University of Cape
Town and SISSA for hospitality while part of this work was carried
out. CGT was partly supported by PPARC (at Portsmouth) and by a
Sida/NRF grant (at Cape Town).

\label{lastpage}


\begin{thebibliography}{99}
\bibitem[Barrow \& Gotz 1989]{BG} Barrow, J.D. and Gotz, G., 1989,
Class.\ Quantum Grav.\ {\bf 6}, 1253

\bibitem[Barrow \& Subramanian 1998]{BS} Barrow, J.D. and
Subramanian, K., 1998, Phys. Rev. D {\bf 58}, 083502

\bibitem[Bruni 1996]{B} Bruni, M., 1996, in Mapping, Measuring and
Modeling the Universe, ASP Conference Series, eds. P. Coles, V.J
Martinez, M.J. Pons-Borderia, Vol. 94

\bibitem[Buchert 1996]{buchert} Buchert, T., 1996, in Proc.
International School of Physics Enrico Fermi, Course CXXXII:
``Dark Matter in the Universe", Varenna 1995, eds. S. Bonometto,
J. Primack, A. Provenzale ( IOP Press, Amsterdam)

\bibitem[Dolag et al 1999]{DBL1} Dolag, K., Bartelmann, M. and
Lesch, H., 1999, A \& A, {\bf 348}, 351

\bibitem[Dolag et al 2002]{DBL2} Dolag, K., Bartelmann, M. and
Lesch, H., 2002, A \& A, {\bf 387}, 383

\bibitem[Ellis 1971]{E} Ellis, G.F.R., 1971, in General Relativity
and Cosmology, ed. R.K. Sachs, Academic, NY, p. 104

\bibitem[Ellis \& Tsagas 2002]{ET} Ellis, G.F.R. and Tsagas, C.G., 2002
(preprint astro-ph/0209143)

\bibitem[Evans \& Fennelly 1985]{EF} Evans, C.R. and Fennelly,
A.J., 1985, Astrophys. Space Sci., {\bf 109}, 15

\bibitem[Grasso \& Rubinstein 2001]{GR} Grasso, D. and
Rubinstein, H.R., 2001, Phys. Rep. {\bf 348}, 163

\bibitem[Han \& Wielebinski 2002]{HW} Han, J-L. and Wielebinski, R.,
2002, Chin. J. Astron. Astrophys. {\bf 2}, 293

\bibitem[Kronberg 1994]{K} Kronberg, P.P., 1994, Rep. Prog. Phys.
{\bf 57}, 325

\bibitem[Matarrese 1996a]{mataa} Matarrese, S., 1996a, in Proc.
International School of Physics Enrico Fermi, Course CXXXII:
``Dark Matter in the Universe", Varenna 1995, eds. S. Bonometto,
J. Primack, A. Provenzale ( IOP Press, Amsterdam)

\bibitem[Matarrese 1996b]{matab}
Matarrese, S., 1996b, in The Universe at High-z, Large Scale
Structure and the Cosmic Microwave Background, eds. E.
Martinez-Gonzalez, J.L. Sanz. Lecture Notes in Physics
(Springer-Verlag, Berlin)

\bibitem[Nusser et al 1991]{nusser} Nusser, A., Dekel., A.,
Bertshinger, E. and Blumenthal, G.R., 1991, Ap. J. {\bf 379}, 6

\bibitem[Padmanabhan 1993]{P} Padmanabhan, T., 1993,
Structure Formation in the Universe (Cambridge Univeristy Press,
Cambridge)

\bibitem[Papadopoulos \& Esposito 1982]{PE} Papadopoulos, D. and
Esposito F.P., 1982, Astrophys. J. {\bf 257}, 10

\bibitem[Roettiger et al 1999]{RSB} Roettiger, K., Stone, J.M. and
Burns, J.O., 1999, Astrophys. J., {\bf 518}, 594

\bibitem[Ruzmaikina \& Ruzmaikin 1971]{RR} Ruzmaikina, T.V. and
Ruzmaikin A.A., 1971, Soviet Astron., {\bf 14}, 963

\bibitem[Shandarin \& Zel'dovich 1989]{SZ} Shandarin, S.F. and
Zel'dovich, Ya.B., 1989, Rev. Mod. Phys. {\bf 61}, 185

\bibitem[Tsagas \& Barrow 1997]{TB} Tsagas, C.G. and Barrow, J.D.,
1997, Class. Quantum Grav. {\bf 14}, 2539

\bibitem[Tsagas \& Maartens 2000]{TM} Tsagas, C.G. and Maartens,
R., 2000, Phys. Rev. D {\bf 61}, 083519

\bibitem[Wasserman 1978]{W} Wasserman, I., 1978, Ap. J. {\bf
224}, 337

\bibitem[Widrow 2002]{Wid} Widrow, L.M., 2002, Rev. Mod. Phys.,
{\bf 74}, 775

\bibitem[Zel'dovich 1970]{Z} Zel'dovich, Y.B., 1970, Astrophysica
{\bf 6}, 319

\bibitem[Zel'dovich et al 1983]{ZRS} Zel'dovich, Ya.B.,
Ruzmaikin, A.A. and Sokoloff, D.D., 1983, Magnetic Fields in
Astrophysics, Gordon and Breach, NY


\end{thebibliography}
\end{document}